\documentstyle{article}\begin{document}
\title{Temperature fluctuations in an inhomogeneous  diffusive fluid }
\author{ Z. Haba\\
Institute of Theoretical Physics, University of Wroclaw,\\ 50-204
Wroclaw, Plac Maxa Borna 9, Poland\\
email:zhab@ift.uni.wroc.pl}\maketitle
\begin{abstract} We discuss metric perturbations of the relativistic diffusion
equation around the homogeneous J\"uttner  equilibrium of massless
particles in a homogenous expanding universe. We describe the
perturbations corresponding to the gravitational wave background.
We show that the lowest order perturbation can be treated as a
variation of temperature. We derive a formula expressing
temperature fluctuations in terms of the diffusion and tensor
power spectrum. We discuss the multipole expansion of the
fluctuations in the presence of diffusion.\end{abstract}
\section{Introduction}The currently accepted concordance cosmological
model ($\Lambda$CDM)\cite{con} is in excellent agreement with
observational data. However, the incorporation of dark matter and
dark energy in this model means that some forms of matter and
their interactions remain unknown. We may assume that at present
only the gravitational interactions are relevant. However, the
unknown particles and their non-gravitational interactions might
play a significant role in the early universe. If the early
universe is described in terms of known particles and interactions
then the lack of information about the remaining forms of matter
will be experienced as a non-conservation of energy-momentum of
the observed particles. The energy will dissipate into a
surrounding of "dark matter" like the energy of Brownian particles
does. We suggest the relativistic diffusion as a model of energy
dissipation. The diffusion comes from a Markovian (no memory)
approximation and does not depend on the details of the
interactions. In our earlier paper \cite{habacqg} we have
discussed a diffusion of massless particles in a homogeneous
expanding metric. The diffusion equation determines a phase-space
distribution which defines the energy-momentum of diffusing
particles. The diffusion equation for massless particles has as a
solution the J\"uttner distribution with a time-dependent
temperature. We have discussed also a non-linear version of the
diffusion equation (taking into account the quantum statistics)
which leads to the time-dependent Planck distribution. We insert
the energy-momentum of diffusing particles into  the Einstein
equations. It comes out that  a compensation of the
energy-momentum non-conservation leads to a time-dependent
cosmological term.  We have shown that the time-dependent
J\"uttner and Planck distributions give the standard Friedmann
equations. We have derived a modified Friedmann equation taking
into account the diffusive energy dissipation. For general
solutions of the diffusion equation the modification leads to a
substantial departure from the Friedmann evolution.

In this letter we report on our  study of diffusing particles
moving in a non-homogeneous metric. We are interested in structure
formation and temperature fluctuations of a dissipative system of
particles. Dissipation is unavoidable in a description of the
evolution of density perturbations and temperature fluctuations in
a surrounding of an unknown matter and interactions. A dissipation
is also necessary in order to achieve a smooth behaviour for a
large time. The deterministic (Hamiltonian) systems experience a
sensitive dependence on initial conditions. Such a dependence is
especially undesirable in cosmological models when the knowledge
of initial conditions is not available. In this paper we continue
a description of dissipation by a relativistic noise. We consider
inhomogeneous metric and discuss a solution of the diffusion
equation as a perturbation in the inverse temperature. We show
that such a perturbative solution of the inhomogeneous diffusion
equation can be treated as a variation of temperature. The
solution of the diffusion equation can be considered as a
perturbation by diffusion of the well-known perturbative solution
of the Liouville-Vlasov equation discussed in the context of the
integrated Sachs-Wolfe effect
\cite{sachs}\cite{sachs2}\cite{mukhanov}. In this preliminary
study we do not intend to explore the complete system of Einstein
equations with diffusing particles. We restrict ourselves to the
lowest order approximation of tensor perturbations. In this
approximation the tensor perturbations can be identified with
gravitational waves.
 The recent
BICEP2 results \cite{bicep} suggest that in an early universe the
perturbations by gravitational waves are of the same order of
magnitude as the curvature perturbations leading to structure
formation. We show that the diffusion can have a significant
impact on temperature fluctuations only if the integrated
Sachs-Wolfe effect is numerically relevant. It has not been
detected in CMB measurements (it is decreasing at small angular
scales and its contribution is more difficult to extract from
experimental data). Nevertheless, with the growing precision of
measurements the integrated Sachs-Wolfe effect corrected by
diffusion could give additional important information about the
physical origins of the departure from homogeneity.

The plan of the paper is as follows. In sec.2 we derive the
well-known \cite{mukhanov}\cite{durrer} perturbative solution of
the Liouville-Vlasov equation for massless particles in an
inhomogeneous metric in a way which can easily be generalized  to
the diffusion equation. In sec.3 we discuss the relativistic
diffusion equation on a space-time dependent metric. The J\"uttner
distribution with a time-dependent temperature is a solution of
the diffusion equation in a homogeneous metric. We look for a
perturbative solution of the diffusion equation in an
inhomogeneous metric  around the J\"uttner distribution. We obtain
a solution which can be interpreted as a diffusive variation of
temperature in a space-time dependent  metric. In sec.4 we
calculate correlation functions of temperature fluctuations in a
diffusive system which result from tensor perturbations in a
quantum vacuum.
 We discuss the form of the temperature correlations and their
 multipole expansion which follow from
  the metric fluctuations. Some examples of metric
fluctuations are discussed in the Appendix.

\section{Einstein-Liouville-Vlasov equations}
In this section we solve perturbatively the
Einstein-Liouville-Vlasov equation describing a distribution of
classical trajectories (see \cite{ellis}\cite{andrea} for its
application in general relativity). We decompose
\begin{equation}g_{\mu\nu}=\overline{h}_{\mu\nu}+h_{\mu\nu},
\end{equation}where
$\overline{h}_{\mu\nu}$ describes homogenous metric in the
conformal time and
\begin{equation}
ds^{2}=g_{\mu\nu}dx^{\mu}dx^{\nu}=a^{2}(dt^{2}-d{\bf
x}^{2}-\gamma_{ij}dx^{i}dx^{j}).
\end{equation}
We decompose the Christoffel symbols corresponding to the
perturbation (1)
\begin{equation}\begin{array}{l}
\Gamma^{\mu}_{\nu\rho}=\overline{\Gamma}^{\mu}_{\nu\rho}+\delta\Gamma^{\mu}_{\nu\rho}.
\end{array}\end{equation}

 Using eq.(3) we write the Liouville equation in the form
\begin{equation}
\begin{array}{l}
(p^{\mu}\partial^{x}_{\mu}-\overline{\Gamma}^{k}_{\mu\nu}p^{\mu}p^{\nu}\partial_{k})\Omega=
\delta\Gamma^{k}_{\mu\nu}p^{\mu}p^{\nu}\partial_{k}\Omega,
\end{array}\end{equation}where
 $\partial_{\mu}^{x}=\frac{\partial}{\partial x^{\mu}}$ and
$x=(t,{\bf x})$ (boldface letters denoting the three vectors),
$\partial_{k}=\frac{\partial}{\partial p^{k}}$ denotes derivatives
over momenta.

For massless particles  and in the homogeneous metric
($h_{\mu\nu}=0$) any function $F(a^{2}\vert{\bf p}\vert)$ is the
solution of eq.(4)\cite{bernstein}. We are interested in classical
equilibrium states described by the J\"uttner distribution
\cite{juttner}
\begin{equation} \Omega_{E}=\exp(-a^{2}\beta\vert{\bf p}\vert)
\end{equation}where
$\beta\equiv\frac{1}{T}$ is   the inverse temperature
and\begin{displaymath} {\bf p}^{2}=\sum_{j}p^{j}p^{j}
\end{displaymath}

In an inhomogeneous metric (2) we consider the J\"uttner
distribution of  the form
\begin{equation} \Omega_{E}^{g}=\exp(-\beta p_{0}),
\end{equation}
where (from $g_{\mu\nu}p^{\mu}p^{\nu}=0$)
\begin{equation}
p_{0}=g_{00}p^{0}=a^{2}({\bf
p}^{2}+p^{j}\gamma_{jk}p^{k})^{\frac{1}{2}}.
\end{equation}
Then,  the solution  of eq.(4) (neglecting second order terms in
$\gamma$) can be expressed as
\begin{equation}
\Omega=\Omega^{g}_{E} +\beta p_{0}\Theta\Omega_{E}^{g},
\end{equation}
 where $n^{k}=p^{k}\vert{\bf p}\vert^{-1}$ and $\Theta $ is the solution of the equation
\begin{equation}\partial_{t}\Theta+n^{k}\partial_{k}^{x}\Theta=-\frac{1}{2}n^{j}n^{k}\partial_{t}\gamma_{jk}.
\end{equation}
 It follows that (till the first order in $\delta T$)
\begin{equation}
\Omega=\exp(-\frac{p_{0}}{T+\delta T}), \end{equation}where
\begin{equation}
\delta T=T\Theta.
\end{equation}

 Eq.(9)
has the solution
\begin{equation}
\begin{array}{l}\Theta(t,{\bf x})=\Theta_{0}({\bf x}-{\bf n} t)-
\frac{1}{2}\int_{0}^{t}\partial_{s}\gamma_{jk}(s,{\bf x}-(t-s){\bf
n})n^{j}n^{k}ds
\end{array}\end{equation}
with the initial condition $\Theta_{0}({\bf x})$. For a realistic
system of a gas of interacting particles the collision terms
should be taken into account. A solution of the  Boltzmann
equation (with collision terms) for the temperature variation in
an inhomogeneous metric has been developed by means of numerical
methods \cite{cmb1}-\cite{cmb2} (CMBFast and CAMB, see also the
discussion in \cite{hu}). In analytic methods some approximations
to the collision term are applied . Dodelson ( sec.4.4 in
\cite{durrer}) derives a friction term as the relaxation time
approximation to the Boltzmann equation (see \cite{kremmer} for a
discussion of such a term in a general metric). In the next
section we consider diffusion approximations which can simulate
interactions with the dark matter (a diffusion approximation for
the collision term in the relativistic Boltzmann equation has been
discussed in \cite{kremmer2}). The diffusion and friction relax
the dependence on the initial conditions (in the sense that the
initial condition is irrelevant for the large time behaviour).
\section{Relativistic  diffusion}
 In this section we first review the theory of the relativistic diffusion. Then, we
 solve the diffusion equation in a space-time dependent metric in a close analogy to the
 solution of the Liouville equation in sec.2. We look
for a relativistic generalization of the Kramers diffusion
defined on the phase space. It is determined in the unique way by
the requirement that the diffusing particle moves on the
mass-shell  ${\cal H}_{+}$  (see
\cite{schay}\cite{dudley}\cite{habapre}\cite{habacqg0}\cite{calo})

\begin{equation}
 g_{\mu\nu}p^{\mu}p^{\nu}=m^{2}.
 \end{equation}
 The diffusion is generated by the Laplace-Beltrami operator $\triangle_{H}^{m}$ on ${\cal H}_{+}$
\begin{equation}
\triangle_{H}^{m}=\frac{1}{\sqrt{G}}\partial_{j}G^{jk}\sqrt{G}\partial_{k},
\end{equation}
where
\begin{equation}
G^{jk}=m^{2}g^{jk}+p^{ j}p^{ k},
\end{equation}  $\partial_{j}=\frac{\partial}{\partial
p^{ j}}$ and $G=\det(G_{jk})$ is the determinant of  $G_{jk}$.

The transport equation for the linear diffusion generated by
$\triangle_{H}$ reads
\begin{equation}
\begin{array}{l}
(p^{\mu}\partial^{x}_{\mu}-\Gamma^{k}_{\mu\nu}p^{\mu}p^{\nu}\partial_{k})\Omega=
\kappa^{2}\triangle^{m}_{H}\Omega,
\end{array}\end{equation} where $\kappa^{2}$ is the diffusion constant.
Eq.(16) is a generalization of the Liouville equation (4)
incorporating dissipation.

Next, we consider   a diffusion at finite temperature
$\beta^{-1}$\cite{habapre}\cite{habacqg0}\cite{calo}\cite{dunkel}\cite{chev}(in
a frame moving with the heat bath; this equation can be considered
as a linear version of the Kompaneets equation
\cite{komp}\cite{habaphysa})

\begin{equation}
\begin{array}{l}
(p^{\mu}\partial^{x}_{\mu}-\Gamma^{k}_{\mu\nu}p^{\mu}p^{\nu}\partial_{k})\Omega=
\kappa^{2}p_{0}\partial_{j}(G^{jk}p_{ 0}^{-1}\partial_{k}+\beta
p^{j})\Omega.
\end{array}\end{equation} Here, $p_{0}$
is determined from eq.(13).
 $\Omega_{E}$ (5) is the  solution of eq.(17) in the
 homogeneous metric in the massless (ultrarelativistic) limit $m=0$.
In our interpretation the expanding dark matter plays the role of
a heat bath. The general solution of eq.(17) tends to $\Omega_{E}$
for a large time.

 In the inhomogeneous metric (2)
 we write the solution in the form (10)
\begin{equation}
\Omega=\Omega_{E}^{g}+\beta p_{0}\Theta\Omega_{E}^{g}+\beta
ra^{2}\Omega_{E}^{g}=(1+\beta
ra^{2})\exp(-\frac{p_{0}}{T+T\Theta})=\exp(-\frac{p_{0}-a^{2}r}{T+T\Theta})
\end{equation}where the equality is up to the second order terms in
$\Theta$ and $r$. It can be seen from eq.(18) that the term
$-a^{2}r$ has the meaning of a supplementary gravitational
potential induced by the diffusion in a heat bath . Inserting this
formula in eq.(17) we obtain equations for the temperature
fluctuation $\Theta$ and  for $r$
\begin{equation}\partial_{t}\Theta+n^{k}\partial_{k}^{x}\Theta+\kappa^{2}\beta a^{2}\Theta=
-\frac{1}{2}n^{j}n^{k}\partial_{t}\gamma_{jk},
\end{equation}
\begin{equation}\partial_{t}r+n^{k}\partial_{k}^{x}r+2{\cal H}r=3\kappa^{2}\Theta,
\end{equation}
where
\begin{equation}
n^{k}=p^{k}\vert{\bf p}\vert^{-1}.
\end{equation}
The solution of eq.(19) reads
\begin{equation}\begin{array}{l}
\Theta_{t}({\bf
x})=\exp(-\beta\kappa^{2}\int_{0}^{t}a^{2}(s)ds)\Theta_{0}({\bf
x}-t{\bf
n})\cr-\frac{1}{2}\int_{0}^{t}ds\exp(-\beta\kappa^{2}\int_{s}^{t}a^{2}(r)dr)\partial_{s}\gamma_{jk}n^{j}n^{k}(s,{\bf
x}-(t-s){\bf n}),\end{array}\end{equation} where $\Theta_{0}$ is
the initial condition. Note that in cosmological models (for a
large $t$ the scale factor $a$ is increasing) the term depending
on the initial condition disappears.  Eq.(22) describes a
variation of the temperature of a gas of particles interacting
with an environment. We have approximated the interaction by
diffusion.

\section{Temperature fluctuations}
The aim of this section is to obtain explicit analytic expressions
for temperature fluctuations. We restrict ourselves to tensor
perturbations \cite{staro}\cite{ruba}(the general perturbations
require a solution of  Einstein equations with an energy-momentum
tensor describing matter distribution \cite{brand}\cite{bardeen}).
The formula for the correlations of temperature fluctuations
follows from eq.(22) and the discussion in Appendix (we neglect
the term depending on the initial condition $\Theta_{0}$ which
disappears at large time)
\begin{equation}
\begin{array}{l}\langle \Theta(t,{\bf n})\Theta(t,{\bf n}^{\prime})\rangle=
 \cr=\frac{1}{4}
(2\pi)^{-3}\int_{0}^{t}ds\int_{0}^{t}ds^{\prime}\int d{\bf
q}F(s,s^{\prime},q)\exp(-\beta\kappa^{2}(\int_{s}^{t}+\int_{s^{\prime}}^{t})dra^{2}(r))\cr(2({\bf
n}\Delta({\bf q}){\bf n}^{\prime})^{2}-({\bf n}\Delta({\bf q}){\bf
n})({\bf n}^{\prime}\Delta({\bf q}){\bf n}^{\prime}))
\exp(-i(t-s){\bf nq}+i(t-s^{\prime}){\bf n}^{\prime}{\bf q}),
\end{array}\end{equation}
where\begin{equation} {\bf n}\Delta({\bf q}){\bf n}^{\prime}={\bf
n}{\bf n}^{\prime}-{\bf q}^{-2}({\bf qn})({\bf
qn}^{\prime})\equiv\Delta({\bf nn}^{\prime},{\bf en},{\bf en}^{\prime})
\end{equation}
\begin{equation}
{\bf n}\Delta({\bf q}){\bf n}=1-{\bf q}^{-2}({\bf qn})^{2}\equiv
\delta({\bf en})
\end{equation}and we write ${\bf q}=q{\bf e}$ with a unit vector ${\bf e}$. In eq.(23)
\begin{equation}
F(s,s^{\prime},q)=\partial_{s}\partial_{s^{\prime}}P(s,s^{\prime},q)\end{equation}
where $P$ is the correlation function (eqs.(2)-(4) in the
Appendix) of tensor perturbations. $ F(s,s,q)$ is the power
spectrum of the gravitational waves (eq.(7) of the Appendix). The
various factors in the formula (23) for temperature fluctuations
result from general rules: the assumption that the tensor
perturbations are transverse and traceless  and the power spectrum
$F$ describes the density of Fourier modes. The extra damping
factor $\exp(-\beta\kappa^{2}\int a^{2})$ comes from the friction
leading to the equilibrium in eq.(17) (a consequence of the
detailed balance). The expression (23) for temperature
fluctuations can be finite for $t\rightarrow\infty$  owing to the
damping exponential factor (even if the corresponding expression
without the diffusion grows
 to infinity).  As pointed out in the Appendix the power spectrum
$F(s,s,q)$ can be associated with the energy of gravitational
waves. By means of the temperature fluctuations (23) $F$ is
accessible to measurements.  We discuss here some consequences of
various forms of $F$.

At the initial stage of our study of eq.(23) we follow the
monograph by Mukhanov \cite{mukhanov}. In the integral $d{\bf
q}=dqq^{2}d{\bf e}$ we  integrate first over ${\bf e}$ in the
exponential in eq.(23). We obtain
\begin{equation} \begin{array}{l}\int d{\bf e}\exp(-i(t-s){\bf nq}+i(t-s^{\prime}){\bf n}^{\prime}{\bf q})\cr=2\pi
q^{-1}\vert (t-s){\bf n}-(t-s^{\prime}) {\bf n}^{\prime}\vert^{-1}
 \sin\Big(\vert(t-s)q{\bf
n}-(t-s^{\prime})q{\bf n}^{\prime}\vert\Big).
\end{array}\end{equation}Next, we use the expansion
\begin{equation}\begin{array}{l}
q^{-1}\vert(t-s){\bf n}-(t-s^{\prime}){\bf n}^{\prime}\vert^{-1}
 \sin\Big(\vert(t-s)q{\bf
n}-(t-s^{\prime})q{\bf n}^{\prime}\vert\Big) \cr
=\sum_{l=0}^{\infty}(2l+1)j_{l}(q(t-s))j_{l}(q(t-s^{\prime}))P_{l}({\bf
n}{\bf n}^{\prime}).
\end{array}\end{equation}  $j_{l}$ is the Bessel spherical
function related to the Bessel function $J$ \cite{grad}
\begin{displaymath}
j_{l}(z)=\sqrt{\frac{\pi}{2z}}J_{l+\frac{1}{2}}(z)
\end{displaymath}
and $P_{l}$ are the Legendre polynomials.

If  $F$ is known then there remains to perform the integrals over
$s$ and $q$ using eqs.(23) and (28) in order to obtain
\begin{equation}
\begin{array}{l}\langle \Theta(t,{\bf n})\Theta(t,{\bf n}^{\prime})\rangle=
 \sum_{l=0}^{\infty}(2l+1)\tilde{D}_{l}(t,{\bf nn}^{\prime})P_{l}({\bf
n}{\bf n}^{\prime})=\sum_{l=0}^{\infty}(2l+1)C_{l}(t)P_{l}({\bf
n}{\bf n}^{\prime}),
\end{array}\end{equation}
In eq.(29) $\tilde{D}_{l} P_{l}$ still must be expanded in
Legendre polynomials if the coefficients $C_{l}$ are to be
independent of the angle. We have from
eqs.(23)-(28)\begin{equation}
\begin{array}{l}\tilde{D}_{l}=
\frac{1}{16\pi^{2}}\int_{0}^{t}ds\int_{0}^{t}ds^{\prime}\int
dqq^{2}F(s,s^{\prime},q)\exp(-\beta\kappa^{2}(\int_{s}^{t}+\int_{s^{\prime}}^{t})dra^{2}(r))\cr\Big(
2\Delta({\bf
nn}^{\prime},-i\partial_{s},i\partial_{s^{\prime}})^{2}-\delta(-i\partial_{s}
)\delta(i\partial_{s^{\prime}})\Big)
j_{l}(q(t-s))j_{l}(q(t-s^{\prime})).
\end{array}\end{equation} Let us
consider only the first term in eq.(30) (denoted $D_{l}$)
resulting from the expansion\begin{displaymath} 2\Delta({\bf
nn}^{\prime},-i\partial_{s},i\partial_{s^{\prime}})^{2}-\delta(-i\partial_{s}
)\delta(i\partial_{s^{\prime}})=2({\bf n
n}^{\prime})^{2}-1+O(\partial_{s},\partial_{s^{\prime}})
\end{displaymath} where $O$ is a polynomial of at least first
order in derivatives. The remaining terms in eq.(30) can easily be
calculated by taking derivatives $\partial_{s}$ of $D_{l}$ ( the
method to express the remaining terms by time derivatives of $D_{l}$
is taken from \cite{mukhanov}). We have
\begin{equation}\begin{array}{l}
D_{l}=\frac{1}{16\pi^{2}} (2({\bf n
n}^{\prime})^{2}-1)\int_{0}^{t}ds^{\prime}\int_{0}^{t}ds
\exp(-\beta\kappa^{2}(\int_{s}^{t}+\int_{s^{\prime}}^{t})dra^{2}(r))\cr
\int_{0}^{\infty}dqq^{2}F(s,s^{\prime},q)j_{l}(q(t-s))j_{l}(q(t-s^{\prime}))
\end{array}\end{equation}
$D_{l}$ as well as $\tilde{D}_{l}$ have the  same behaviour at
large $l$ as $C_{l}$ in eq.(29).
 We make  a
simplifying assumption
\begin{equation}
F(s,s^{\prime},q)=f(s,s^{\prime})\sigma(q).
\end{equation}First, we estimate $D_{l}$ for a large $l$. According to
the Limber asymptotic formula  \cite{limber1}\cite{limber2} (see
also the calculations in sec.9 of \cite{rubakov2})
\begin{equation}\begin{array}{l}
D_{l}=\frac{1}{32\pi} (2({\bf n
n}^{\prime})^{2}-1)\int_{0}^{t}ds(t-s)^{-2} \cr
f(s,s)\exp\Big(-2\beta\kappa^{2}\int_{s}^{t}dra^{2}(r)\Big)
\sigma(\frac{l+\frac{1}{2}}{t-s})\Big(1+O(l^{-2})\Big).
\end{array}\end{equation}Here, $O(l^{-2})$ denotes terms of order $l^{-2}$.
For the inflationary  almost flat  spectrum (small $\epsilon$)
 \begin{equation} \sigma(q)=Aq^{-3+\epsilon}.\end{equation}
Eq.(34) can also apply to the model describing the appearance of
inflation after the equilibrium phase \cite{indian}. Then,
$\epsilon\rightarrow -1+\epsilon$ in eq.(34). For any $\epsilon<2$
in the power spectrum (34)
 we get from eq.(33)the asymptotic behaviour
 \begin{equation}\begin{array}{l}
D_{l}=A(l+\frac{1}{2})^{-3+\epsilon}\frac{1}{32\pi} (2({\bf n
n}^{\prime})^{2}-1)\int_{0}^{t}ds(t-s)^{1-\epsilon}\cr
f(s,s)\exp\Big(-2\beta\kappa^{2}\int_{s}^{t}dra^{2}(r)\Big)
(1+O(l^{-2})).
\end{array}\end{equation}As can be seen from eq.(35) the asymptotics in $l$ does not depend on the diffusion.
We can derive an exact result for the integral (31) using the
formula 6.574 of \cite{grad}($\gamma<\sigma$)
\begin{equation}
\begin{array}{l}
\int_{0}^{\infty}dqq^{-2+\epsilon}J_{l+\frac{1}{2}}(\sigma
q)J_{l+\frac{1}{2}}(\gamma q)
=\frac{1}{2^{2-\epsilon}}\frac{\Gamma(l+\frac{\epsilon}{2})}{\Gamma(\frac{3}{2}-\frac{\epsilon}{2})\Gamma(l+\frac{3}{2})}
(\frac{\gamma}{\sigma})^{l}\sqrt{\gamma\sigma}\sigma^{\epsilon}\cr
F(l+\frac{\epsilon}{2},-\frac{1}{2}+\frac{\epsilon}{2},l+\frac{3}{2},\frac{\gamma^{2}}{\sigma^{2}}),
\end{array}\end{equation}where $F(\alpha,\beta,\gamma,z)$ denotes
the hypergeometric function. Applying eq.(36) we
obtain\begin{equation}\begin{array}{l} D_{l}(t)=\frac{1}{16\pi}
A(2({\bf n
n}^{\prime})^{2}-1)\frac{1}{2^{2-\epsilon}}\frac{\Gamma(l+\frac{\epsilon}{2})}{\Gamma(\frac{3}{2}-\frac{\epsilon}{2})
\Gamma(l+\frac{3}{2})} \int_{0}^{t}ds\int_{0}^{s}ds^{\prime}
(\frac{t-s^{\prime}}{t-s})^{l} (t-s)^{\epsilon}\cr
f(s,s^{\prime})\exp\Big(-\beta\kappa^{2}(\int_{s}^{t}+\int_{s^{\prime}}^{t})dra^{2}(r)\Big)
 F(l+\frac{\epsilon}{2},-\frac{1}{2}+\frac{\epsilon}{2},l+\frac{3}{2},\frac{(t-s^{\prime})^{2}}{(t-s)^{2}}).\end{array}\end{equation}
 The integral (31) can easily be calculated if  $F$ (32) is concentrated  at
 $s=s^{\prime}=s_{d}$. This case describes an instantaneous metric perturbation (the metric perturbation is
 limited to the moment $s_{d}$)
 corresponding to a sudden decoupling at $s=s_{d}$ from the last scattering surface \cite{mukhanov}\cite{rubakov2}.
  In such a case we set
 \begin{displaymath}
 f(s,s^{\prime})=f_{d}\delta(s-s_{d})\delta(s^{\prime}-s_{d}).
 \end{displaymath}
 Hence, $s=s^{\prime}=s_{d}$ in the argument of the hypergeometric
 function (36). The metric  perturbation at a fixed time  is applied
 in the ordinary Sachs-Wolfe effect
 \cite{sachs} . In order to relate the integral (31) to the well-known result let
 us calculate it at $s=s^{\prime}$ and $\epsilon=0$ (Harrison-Zeldovich spectrum) in eq.(36).
 Then,  we can obtain the
  value of the hypergeometric function at 1 using the formula
\begin{displaymath}
F(\alpha,\beta,\gamma,1)=(F(\alpha,-\beta,\gamma-\beta,1))^{-1}
\end{displaymath}
and \begin{displaymath} F(\alpha,\beta,\gamma,1)=
\frac{\Gamma(\gamma)\Gamma(\gamma-\alpha-\beta)}{\Gamma(\gamma-\alpha)\Gamma(\gamma-\beta)}.
\end{displaymath}The approximation $\gamma\simeq \sigma=t-s_{d}$ in eq.(37) leads to the result\begin{equation}
\begin{array}{l}
\int_{0}^{\infty}dqq^{-2}J_{l+\frac{1}{2}}(\gamma
q)J_{l+\frac{1}{2}}(\gamma q)
=\frac{\gamma}{\pi}\frac{(l-1)!}{(l+1)!}.
\end{array}\end{equation}As a consequence
\begin{equation}\begin{array}{l}
D_{l}(t)=Af_{d}(2({\bf n
n}^{\prime})^{2}-1)\frac{1}{16\pi}(l(l+1))^{-1}
\exp(-2\beta\kappa^{2}\int_{s_{d}}^{t}dra^{2}(r)).\end{array}\end{equation}
Hence, if  the metric perturbation is applied only at a fixed time
then the diffusion gives just
 a constant factor in the standard formula for the  temperature fluctuations \cite{mukhanov}\cite{rubakov2}.
This behaviour can change when we admit power spectra which are
not of the power-like form . As an example (which can be related
to the Gibbs damping of the high energy modes) we consider the
metric fluctuations of the form \begin{equation}
F(s,s^{\prime},q)=Qq^{n-1}\exp(-\beta q) f(s,s^{\prime}).
\end{equation}
 For the power spectrum (40) we have
the Limber formula

 \begin{equation}\begin{array}{l}
D_{l}(t)=(l+\frac{1}{2})^{n-1}\frac{1}{32\pi} Q(2({\bf n
n}^{\prime})^{2}-1)\cr
\int_{0}^{t}dsf(s,s)(t-s)^{-1-n}\exp\Big(-\frac{\beta(l+\frac{1}{2})}{t-s}-\beta\kappa^{2}\int_{s}^{t}dra^{2}(r)\Big)
 \end{array}\end{equation}
The integral (41) could be calculated by means of the saddle point
method. The saddle point $s_{c}$ is determined from the equation
\begin{equation}
-\frac{1+n}{t-s_{c}}+\beta\frac{l+\frac{1}{2}}{(t-s_{c})^{2}}=\beta\kappa^{2}a^{2}(s_{c})\end{equation}
In eqs.(41)-(42 ) the dependence of $D_{l}$  on $l$ is a function
of the expansion scale factor $a$ and the dissipation rate
$\beta\kappa^{2}$. Note that if $\kappa=0$ then from eq.(41)  we
obtain $D_{l}\simeq l^{-1}$.

In order to derive an exact formula for the power spectrum (40) we
can use the integral 6.612 of \cite{grad}
\begin{equation}\begin{array}{l}
\int_{0}^{\infty}dqq^{n}\exp(-\lambda q)J_{l+\frac{1}{2}}(\sigma
q)J_{l+\frac{1}{2}}(\gamma
q)=\sqrt{\gamma\sigma}(-1)^{n}\frac{l!}{\Gamma(l+\frac{3}{2})}(\gamma\sigma)^{l}
\cr
\partial_{\lambda}^{n}(\lambda^{2}+\gamma^{2}+\sigma^{2})^{-l-1}
F\Big(\frac{1}{2}l+1,\frac{1}{2}l+\frac{1}{2},l+\frac{3}{2},
\frac{4\gamma^{2}\sigma^{2}}{(\lambda^{2}+\gamma^{2}+\sigma^{2})^{2}}\Big).
\end{array}
\end{equation}
The formula (43) can be applied for the calculation of $D_{l}$ by
means of an expansion of the hypergeometric function
 in a power series  because $\lambda>0$ makes the series quickly
convergent. Then,
\begin{equation}\begin{array}{l}
D_{l}(t)=\frac{1}{32\pi}(-1)^{n}Q(2({\bf n n}^{\prime})^{2}-1)
\int_{0}^{t}ds^{\prime}\int_{0}^{s^{\prime}}ds
(\frac{t-s^{\prime}}{t-s})^{l} \cr
\exp\Big(-\beta\kappa^{2}(\int_{s}^{t}+\int_{s^{\prime}}^{t})dra^{2}(r)\Big)\partial_{\lambda}^{n}
F(\frac{1}{2}l+1,\frac{1}{2}l+\frac{1}{2},l+\frac{3}{2},\frac{4\gamma^{2}\sigma^{2}}{(\lambda^{2}+\gamma^{2}+\sigma^{2})^{2}}),
\end{array}
\end{equation}
where $\gamma=t-s$ and $\sigma=t-s^{\prime}$. The analytic
expressions derived in this section give an explicit large $l$
behaviour. The integral formulae  for fluctuations could be
applied for numerical integration . In comparison with the general
Boltzmann equations \cite{cmb1}\cite{cmb2} our expressions are
rather elementary because we replace the complex collision
integrals by the relativistic diffusion.

 \section{Discussion and Outlook}
 We have found solutions of the relativistic diffusion equation which can be considered as  a diffusive
 disturbance  of the standard Liouville equation in
 a perturbed metric. The diffusion can approximate the collision term in the Boltzmann equation
  when an interaction with the dark matter is taken into account. The solutions
 describe temperature variation $\Theta$ in an inhomogeneous metric.
 In order to determine the temperature variation unambiguously  for general perturbations we would need
 to solve the Einstein equations with the energy-momentum tensor
 defined by temperature fluctuations. The energy-momentum  is determined by $\Theta$ and is of the form $\simeq
 (T+T\Theta)^{4}$. In this paper we restrict ourselves to tensor
 perturbations which in the lowest order (i.e. , with no energy-momentum
 on the rhs of Einstein equations) are identical with
 gravitational waves. Their quantization leads to temperature
 fluctuations. The recent BICEP2 results indicate that such temperature
 fluctuations are of the same order of magnitude as the fluctuations
 resulting from scalar perturbations. Our results for temperature variation
 reduce to the ones derived earlier by other authors in the absence of diffusion.
 In the sudden decoupling approximation (ordinary Sachs-Wolfe
 effect),
 confirmed by observations within the present experimental accuracy, the dissipation
 gives
 only a multiplicative constant. However, in the general formula
 for the integrated Sachs-Wolfe effect the diffusion contributes
 along the whole particle's path .
 It  can give an $l$-dependent modification to
 the multipole expansion coefficients $C_{l}$.
 The integrated Sachs-Wolfe effect is  negligible at large $l$ in
 comparison with the ordinary Sachs-Wolfe effect. At the
 present precision of measurements it seems unlikely that the
 effect of diffusion  on tensor perturbations can be measurable. Nevertheless, a determination
 of the  contribution of diffusion may suggest future measurements.
 We intend to continue the study of the impact of diffusion upon remaining
 perturbations of the metric (especially scalar perturbations leading to acoustic oscillations).
 For this purpose we must investigate
 Einstein equations for the perturbations.

{\bf Acknowledgement}

The research is supported by NCN grant DEC-2013/09/B/ST2/03455

\section{Appendix}\setcounter{equation}{0}In general, we should
insert the energy-momentum tensor $T^{\mu\nu} $ determined by the
diffusion (17) into the Einstein equations in order to derive the
solution for the metric perturbations
$h_{\mu\nu}=a^{2}\gamma_{\mu\nu}$. The solution depends on the
linear gravitational waves which are the solutions of Einstein
equations at the zeroth order of perturbation corresponding to
$T^{\mu\nu}=0$. Inserting the solutions $\gamma_{jk}$  in $\Theta$
(22) we calculate the temperature fluctuations resulting from
gravitational waves in the lowest order of perturbation.

 We expand  solutions of Einstein equations
in an external  homogeneous metric in plane waves
\begin{equation}
\gamma_{jk}(t,{\bf x})=(2\pi)^{-\frac{3}{2}}\sum_{\lambda}\int
d{\bf q}\vert {\bf q}\vert^{-\frac{1}{2}}(a(\lambda,{\bf
q})e_{jk}(\lambda,{\bf q})\gamma(t,{\bf q},\lambda)\exp(i{\bf
qx})+cc),
\end{equation} where $cc$ denotes the complex conjugation of the preceding term,
$\lambda$ is the polarization of gravitons, $e_{jk}$ are the
polarization tensors and $\gamma$ are solutions of the wave
equations ( see \cite{giova}). Quantizing the Fourier modes we
obtain the vacuum correlation functions of gravitons in the
transverse-traceless gauge. So, in the momentum space
\begin{equation}
\langle \gamma_{jn}(t,{\bf k})\gamma_{lr}(t^{\prime},{\bf
q})\rangle= \delta({\bf k}+{\bf q})(2\pi)^{-3}P(t,t^{\prime},q)
(\Delta_{jl}\Delta_{rn}+\Delta_{jr}\Delta_{nl}-\Delta_{jn}\Delta_{lr}),
\end{equation}
where \begin{equation}\Delta_{jl}=\delta_{jl}-q_{j}q_{l}{\bf q}^{-2}
\end{equation} and\begin{equation}
P(t,t^{\prime},q)=q^{-1}\sum_{\lambda,k,j}\overline{e^{k}_{j}(\lambda,{\bf
q})\gamma(t,{\bf q},\lambda)}e^{k}_{j}(\lambda,{\bf
q})\gamma(t^{\prime},{\bf q},\lambda)
\end{equation}$P$ depends only on $q=\vert {\bf q}\vert$ because of the rotational invariance. Taking the time
derivatives in eq.(2)
\begin{equation}
\langle \partial_{t}\gamma_{jn}(t,{\bf
k})\partial_{t^{\prime}}\gamma_{lr}(t^{\prime},{\bf
q}^{\prime})\rangle= \delta({\bf k}+{\bf
q})(2\pi)^{-3}F(t,t^{\prime},{\bf q})
(\Delta_{jl}\Delta_{rn}+\Delta_{jr}\Delta_{nl}-\Delta_{jn}\Delta_{lr}),
\end{equation}
where
\begin{equation} F(t,t^{\prime},{\bf
q})=\partial_{t}\partial_{t^{\prime}}P(t,t^{\prime},{\bf q}).
\end{equation}
The power spectrum $F(t,t,q)$ is  related to the energy
$\epsilon_{g}$ of gravitational waves \cite{arno}
\begin{equation}
\epsilon_{g}=\frac{a^{2}}{32\pi G}\int d{\bf q}
F(t,t,q)\end{equation}

We can understand the power spectrum appearing in temperature
fluctuations in a wider sense as the energy impulse coming from
gravitational waves. We could obtain such generalized $F$
calculating the correlation functions of $\gamma$ in quantum
states more general than just the vacuum states. In such a case
the impulse described by $F$ can be confined in time.

 As a standard example leading to the power spectrum (34) with $\epsilon=0$ we consider
  de Sitter space
with the metric $ds^{2}=t^{-2}(dt^{2}-d{\bf x}^{2})$. The
correlation functions in the Bunch-Davis vacuum
are\begin{equation}P_{dS}(t,t^{\prime},{\bf
q})=\frac{1}{2\vert{\bf
q}\vert}\Big(1+\frac{i(t-t^{\prime})}{\vert{\bf q}\vert
tt^{\prime}}+({\bf
q}^{2}tt^{\prime})^{-1}\Big)\exp(-i(t-t^{\prime})\vert {\bf
q}\vert)\end{equation}For a small $q$
\begin{equation}F(t,t,q)=
\frac{1}{2}q^{-3}t^{-4}\end{equation}


\begin{thebibliography}{99}\bibitem{con}D. S. Gorbunov and V. A. Rubakov, Introduction to the
Theory of the Early Universe. Hot Big Bang Theory,World
Scientific, Singapore, 2011

\bibitem{habacqg}Z. Haba, Class.Quant.Grav.{\bf 31},075011(2014)

\bibitem{sachs}R. Sachs and A. Wolfe, Astroph.J.{\bf 147},73(1967)
\bibitem{sachs2}A. J. Nishizawa, arxiv:1404.5102[astro-ph.CO]

\bibitem{mukhanov}V. Mukhanov, Physical Foundations of Cosmology,


Cambridge Univ.Press,2005\bibitem{bicep}P.A.R. Ade, et al arXiv:1403.3985[astro-ph.CO]

\bibitem{durrer}S. Dodelson, Modern Cosmology, Academic Press,
New York, 2003

\bibitem{ellis}D. Bancel and Y. Choquet-Bruhat ,


 Commun.Math.Phys.{\bf 33},83(1973)


\bibitem{andrea}H. Andreasson, Living Rev.Relativity, {\bf
14},4(2011);arXiv:1106.1367[gr-qc]

\bibitem{bernstein}J. Bernstein, Kinetic Theory in an Expanding Universe,
Cambridge,1988

\bibitem{juttner}F. J\"uttner, Ann.Phys.(Leipzig){\bf 34},856(1911)

\bibitem{cmb1} U. Seljak and M. Zaldariagga, Astroph J. {\bf 469},437(1996)
\bibitem{cmb2}A. Lewis, A. Challinor and A. Lasemby, Astroph.J. {\bf 538} 473(2000)
\bibitem{hu}W. Hu and T. Okamoto, arXiv:03080049[astro-ph.CO]
\bibitem{kremmer}G. M. Kremer, Int.J. Geom.Methods Mod.Phys.{\bf 11},140006(2014)
\bibitem{kremmer2}G. Chacon-Acosta and G.M.Kremer, Phys.Rev.{\bf
E76},021201(2007)

\bibitem{schay} G.Schay,PhD thesis,Princeton University,1961
\bibitem{dudley} R.Dudley, Arkiv for Matematik,{\bf 6},241(1965)
\bibitem{habapre}Z. Haba, Phys.Rev.{\bf E79},021128(2009)
\bibitem{habacqg0}Z. Haba, Class. Quantum Grav.{\bf 27},095021(2010)
\bibitem{calo}J.A. Alcantara and S.Calogero, Kin.Rel. Mod.{\bf
4},401(2011)
\bibitem{dunkel}J. Dunkel and P. H\"anggi, Phys.Rev.{\bf E72},036106(2005)

\bibitem{chev}C. Chevalier and F. Debbasch,

J.Math.Phys.{\bf 49},043303(2008)






\bibitem{komp}A.S.Kompaneets, Sov.Phys.JETP {\bf 4},730(1957)

\bibitem{habaphysa}Z. Haba, Physica{\bf A390},2776(2011)


\bibitem{staro}A. Starobinskii, JETP Lett.{\bf 30},683(1979)
\bibitem{ruba}
V. Rubakov, M. Sazhin and A. Veryaskin, Phys.
Lett.{117B},175(1982)
\bibitem{brand}V.F. Mukhanov, H.A. Feldman and
R.H.Branderberger,

Phys.Rep.{\bf 215},203(1992)
\bibitem{bardeen}J. Bardeen, Phys.Rev.{\bf D22},1882(1980)
\bibitem{limber1}D.N. Limber, Astroph.J.{\bf 117}, 134(1953)
\bibitem{limber2}M. LoVerde and N. Afshordi, Phys. Rev.{\bf
D78},123506(2008)
\bibitem{rubakov2} D.S. Gorbunov and V.A.
Rubakov, Introduction to the Theory of the Early Universe.
Cosmological Perturbations and Inflationary Theory, World
Scientific, 2011
\bibitem{indian} K. Bhattacharya, S. Mohanty and
A. Nautiyal,

Phys. Rev.Lett.{\bf 97},251301(2006)
\bibitem{arno}C.W.Misner, K.S. Thorne and J.A. Wheeler, Gravitation,

W.H. Freeman Co.,San Francisco,1973
\bibitem{grad}Y.S. Gradshtein and Y.M. Ryzhik, Tables of
Integrals, Sums, Series and Products,Nauka, Moscow, 1971 (in
Russian)
\bibitem{giova} M. Giovannini, arXiv:1405.630[astro-ph.CO]




 \end{thebibliography}
\end{document}